\documentclass[conference]{IEEEtran}
\IEEEoverridecommandlockouts
\usepackage{cite}
\usepackage{amsmath,amssymb,amsfonts}
\usepackage{algorithmic}
\usepackage{graphicx}
\usepackage{textcomp}
\usepackage{xcolor}
\def\BibTeX{{\rm B\kern-.05em{\sc i\kern-.025em b}\kern-.08em
    T\kern-.1667em\lower.7ex\hbox{E}\kern-.125emX}}
    
\usepackage{comment}
\usepackage{hyperref}
\usepackage{cleveref}

\usepackage[nonumberlist]{glossaries}
\usepackage{glossary-inline}

\newacronym{CPU}{CPU}{central processing unit}
\newacronym{GPU}{GPU}{graphics processing unit}
\newacronym{QPU}{QPU}{quantum processing unit}
\newacronym{HPC}{HPC}{high performance computing}
\newacronym{LRZ}{LRZ}{Leibniz Supercomputing Centre}
\newacronym{DSL}{DSL}{domain-specific language}
\newacronym{eDSL}{eDSL}{embedded domain-specific language}
\newacronym{NISQ}{NISQ}{noisy intermediate-scale quantum}
\newacronym{MPI}{MPI}{Message Passing Interface}
\newacronym{QRAM}{QRAM}{quantum random access machine}
\newacronym{HQCC}{HQCC}{heterogeneous quantum-classical computation}
\newacronym{API}{API}{application programming interface}
\newacronym{QPT}{QPT}{quantum programming tool}
\newacronym{QC}{QC}{quantum computing}
\newacronym{ISA}{ISA}{instruction set architecture}
\newacronym{RISC}{RISC}{reduced instruction set computer}
\newacronym{RAM}{RAM}{random access memory}
\newacronym{IR}{IR}{intermediate representation}
\newacronym{SQRAM}{SQRAM}{sequential QRAM}
\newacronym{HPCQC}{HPCQC}{high performance computing - quantum computing}
\newacronym{QIR}{QIR}{Quantum Intermediate Representation}
\newacronym{SQIR}{SQIR}{standardized quantum intermediate representation}
\newacronym{AOT}{AOT}{ahead-of-time}
\newacronym{JIT}{JIT}{just-in-time}
\glsdisablehyper 

\usepackage{tikz}

\usepackage[colorinlistoftodos,prependcaption,textsize=tiny,disable]{todonotes} 

\newcommand*{\mycomment}[3][]{%
    \PackageWarning{Unresolved comment}{by #2}
    \todo[inline, size=\small, #1]{\textbf{#2:} #3}
}

\newcommand{\AMR}[2][]{\mycomment[#1, color=orange]{Amr}{#2}}

\newcommand{\RONE}[2][]{\mycomment[#1, color=yellow]{Reviewer One}{#2}}
\newcommand{\RTWO}[2][]{\mycomment[#1, color=lightgray]{Reviewer Two}{#2}}
\newcommand{\RTHREE}[2][]{\mycomment[#1, color=pink]{Reviewer Three}{#2}}

\makeatletter
\newcommand{\linebreakand}{%
  \end{@IEEEauthorhalign}
  \hfill\mbox{}\par
  \mbox{}\hfill\begin{@IEEEauthorhalign}
}
\makeatother

\newcommand\copyrighttext{%
  \footnotesize  © 20xx IEEE. Personal use of this material is permitted. Permission from IEEE must be
obtained for all other uses, in any current or future media, including
reprinting/republishing this material for advertising or promotional purposes, creating new
collective works, for resale or redistribution to servers or lists, or reuse of any copyrighted
component of this work in other works.}
\newcommand\copyrightnotice{%
\begin{tikzpicture}[remember picture,overlay]
\node[anchor=south,yshift=10pt] at (current page.south) {\fbox{\parbox{\dimexpr\textwidth-\fboxsep-\fboxrule\relax}{\copyrighttext}}};
\end{tikzpicture}%
}

\begin{document}

\title{Toward a Unified Hybrid HPCQC Toolchain\\
}

\author{
\IEEEauthorblockN{Philipp Seitz}
\IEEEauthorblockA{\emph{Chair of Scientific Computing in Computer Science} \\
\emph{Technical University of Munich}\\
Munich, Germany \\
philipp.seitz@tum.de}
\and
\IEEEauthorblockN{Amr Elsharkawy}
\IEEEauthorblockA{\emph{Chair of Computer Architecture and Parallel Systems} \\
\emph{Technical University of Munich}\\
Munich, Germany \\
amr.elsharkawy@in.tum.de}
\and
\linebreakand
\IEEEauthorblockN{Xiao-Ting Michelle To}
\IEEEauthorblockA{\emph{MNM Team} \\
\emph{Ludwig-Maximilians-Universität in Munich}\\
Munich, Germany \\
michelle.to@nm.ifi.lmu.de}
\and
\IEEEauthorblockN{Martin Schulz}
\IEEEauthorblockA{\emph{Chair of Computer Architecture and Parallel Systems} \\
\emph{Technical University of Munich}\\
Munich, Germany \\
schulzm@in.tum.de}
}

\maketitle

\begin{abstract}
In the expanding field of \gls{QC}, efficient and seamless integration of \gls{QC} and \gls{HPC} elements (e.g., quantum hardware, classical hardware, and software infrastructure on both sides) plays a crucial role. 
This paper addresses the development of a unified toolchain designed for hybrid quantum-classical systems. 
Our work proposes a design for a unified hybrid \gls{HPCQC} toolchain that tackles pressing issues such as scalability, cross-technology execution, and \gls{AOT} optimization. 
\glsresetall
\end{abstract}
\begin{IEEEkeywords}
Quantum Computing, High Performance Computing, HPCQC Integration.
\end{IEEEkeywords}
\copyrightnotice{}
\section{Introduction}\label{sec:Introduction}

In the rapidly evolving landscape of \gls{QC}, one of the most promising and challenging frontiers is the integration of quantum and classical computing into hybrid systems~\cite{humbleQuantumComputers2021, ruefenacht2022ea, bartsch_2021, Schulz:2022:AHQ}. 
These quantum-classical systems seek to leverage the unique strengths of both paradigms to solve complex problems, such as quantum chemistry~\cite{chemical2020} or cryptography~\cite{cryptography2020}, more efficiently (e.g., faster, using less resources, or with higher precision) than either could independently. 
As this novel computational paradigm emerges, the need for a unified hybrid toolchain that can seamlessly manage and optimize the interplay between quantum and classical \gls{HPC} is more critical than ever~\cite{Hybrid_McCaskey,mccaskey2018language}.

The toolchain in this context refers to a set of interconnected software tools designed to manage the operation of a hybrid quantum-classical system~\cite{ali2012towards,ameen2017towards}. 
Existing toolchains often cater to either quantum or classical systems and exhibit limitations when tasked to handle both simultaneously~\cite{mccaskey_xacc_2020}. 
This limitation forms the problem space this paper addresses.

We introduce a unified hybrid toolchain explicitly designed for hybrid quantum-classical systems. 
Its development is still an ongoing process.
In the future, we plan to refine the design, test the implementation on actual hardware, and integrate \gls{HPCQC} compilation tools.
This toolchain enhances the efficiency and scalability of such systems, providing a more integrated platform to manage the intricate computational dynamics involved. 
Our toolchain stands out by its detailed architecture, designed to manage both quantum and classical computations, and its versatility, which allows the user to adapt it to various quantum hardware and problem contexts.

The paper is structured as follows. 
In \Cref{sec:BackgroundAndRelatedWork}, we introduce the core concepts of \gls{HPCQC} integration and highlight the relevant ongoing research and the shortcomings of the currently available toolchains. 
\Cref{sec:Motivation} contains a use case and describes the guiding principles to our design. 
We present our proposal for a unified \gls{HPCQC} toolchain in \Cref{sec:Toolchain}. 
\Cref{sec:DiscussionandFutureWork} discusses the current state of our work, resulting emerging insights and necessary future research, and we conclude the work in \Cref{sec:Conclusion}.

\section{Background and Related Work}\label{sec:BackgroundAndRelatedWork}


Integrating \gls{HPC} and \gls{QC} on the \emph{software level} relies on creating a robust \emph{software stack}. 
This stack transforms hybrid \gls{HPCQC} algorithms into an executable form on both \gls{HPC} systems and quantum hardware~\cite{ruefenacht2022ea,Schulz:2022:AHQ}.
This seamless interoperability is realized through \glspl{IR}, which are a crucial bridge between hybrid high-level programming languages and low-level machine languages. 
For instance, the \gls{QIR}~\cite{QIRSpec2021} is a recognized LLVM-based~\cite{LLVM_CGO04} intermediary for \gls{HPCQC} programs that underscores the necessity for standardization. 
Such a standard provides a convenient way to support more sophisticated \gls{HPCQC} hardware architectures in data centers as the technology matures. 
In our previous work, we have identified four \gls{HPCQC} hardware integration modes that either are currently available or expected to exist in the future, including loose integration (standalone or co-located) and tight integration (co-located or on-node)~\cite{HPCQC_review_2023}. 





There are already several approaches for a hybrid toolchain. 
One of the first is the LLVM-based compiler ScaffCC~\cite{JAVADIABHARI20152} for the quantum programming language Scaffold~\cite{abhari_scaffold_2012}. 
The compiler separates the program into Scaffold-specific quantum modules represented with classical code, called CTQG, and hybrid quantum modules. 
In CTQG modules, classical code is transformed into quantum gates through a specific compilation procedure, ultimately producing quantum assembly formats like QASM. 
The remaining modules are translated to an LLVM \gls{IR} where classical control instructions are processed. 
Next, the modules are translated to QASM and merged with the CTQG modules in a common QASM variant. 
After gate decomposition, the quantum program analysis checks the resulting program using another LLVM \gls{IR}.
In an extension~\cite{Litteken_2020}, the LLVM \gls{IR} is translated to OpenQASM~\cite{cross_open_2017} to run Scaffold programs on quantum devices.

The most popular toolkit is Qiskit~\cite{qiskit}, implemented in Python.
Thus, the classical components are also run and compiled in Python.
However, it is also possible to define the quantum circuit in OpenQASM~\cite{cross_open_2017,cross_openqasm_2022}.
When users want to execute a quantum circuit, they can simulate it locally or send it to the IBM Quantum cloud platform.
The toolchain represents the circuit internally as a directed acyclic graph.
Before running, the given circuit is transpiled into the native gate set of the quantum device the program should run on.
Finally, the job is scheduled for the specified hardware and runs asynchronously.

QCOR~\cite{mintz_qcor_2019} aims at heterogeneous \gls{HPC} with \glspl{QPU} as a type of accelerator.
It allows user-defined compilation through a visitor pattern, which works with multiple \glspl{IR}, including QIR.
Underneath, QCOR builds on the XACC~\cite{mccaskey_xacc_2020} runtime, which provides a memory model and enables task-level parallelism.
Multiple hardware backends are supported through an accelerator abstraction.

Another example is the toolchain of PennyLane Catalyst~\cite{pennylane_catlyst}:
its infrastructure is designed to allow the compilation of quantum circuits with classical control flow.
A Python program is first translated into MLIR~\cite{mlir}, extended with a quantum dialect, as input for the compiler.
The compiler consists of compilation passes, including custom ones, and the execution follows a defined pass order.
During the compilation, different MLIRs are passed down to LLVM and QIR~\cite{QIRSpec2021} and then output as binary, which the runtime handles.
The Catalyst runtime then links the program to a quantum device for execution.
The toolchain also provides \gls{JIT} quantum compilation: 
it compiles functions marked for \gls{JIT} compilation when called for the first time.
The resulting precompiled binary does not need recompilation; only the parameter values must be inserted.

Intel recently introduced a LLVM-based compiler with custom quantum extension as part of the Intel SDK~\cite{khalate2022llvmbased}.
It is proposed explicitly for implementing hybrid quantum-classical algorithms.
The rough workflow is as follows:
the input is a hybrid quantum-classical C++ source file, which is then processed into a binary executable by the compiler. 
In this compiler toolchain, Intel's quantum-classical \gls{IR} and their own quantum \gls{ISA} is used.
In the next step, the quantum runtime resolves parameters unknown at compile time and updates the corresponding variables.
Then the instructions are translated to control signals, which can be used to run the program on quantum devices or simulators.

NVIDIA has published a similar stack called CUDA Quantum~\cite{cuda_quantum}.
The general concepts are similar to Catalyst:
it is compatible with LLVM and QIR, providing custom dialects and passes for MLIR.
It provides a compiler ``NVQ++'' which compiles quantum kernels to customizable hardware.
Quantum kernels are called from C++ host code (entry-point kernels) or other quantum kernels (pure-device kernels).
CUDA Quantum primarily targets \glspl{GPU}.

This work extends the existing research and developments, proposing a novel, unified \gls{HPCQC} toolchain that enhances the interaction between classical and quantum computational paradigms. Our work is distinguished by the in-depth integration of both computational models, classical and quantum, utilizing emerging standards such as QIR and OpenQASM3.

\section{Guiding Principles}\label{sec:Motivation}

The goal of our proposal is the application in the \gls{HPCQC} domain, which inhibits unique challenges.
Our design choices for the toolchain are motivated in this section.
We evaluate our approach in multiple dimensions derived from the requirements of the setting.
\gls{HPCQC} is subject to many different stakeholders, which translates to multiple perspectives.
Naturally, we focus on the view of software developers, but we strive to consider application engineers, hardware providers, and \gls{HPC} centers themselves as well.

To provide concrete insight, we focus on one generic use case. 
We envision a hybrid quantum chemistry simulation consisting of demanding classical tasks as well as \gls{QC} tasks.
Each task could assemble a quantum subroutine with various hardware requirements.
A problem-specific technique construes the assembly process available as a standalone tool.
The entire program gets deployed to an \gls{HPC} center which offers two interconnected \glspl{QPU}, one integrated \gls{QPU}, and multiple classical nodes.

As mentioned in \Cref{sec:BackgroundAndRelatedWork}, other toolchains already exist.
They fail to capture the full complexity of the described task or come short in certain areas. 
To achieve the required level of complexity, we follow some guiding principles detailed below.

\subsection{Performance} \label{subsec:Performance}

Performance is a loosely defined concept and depends on the context.
From a user's perspective, time-to-solution is critical, as computation time is expensive.
Multiple criteria exist for the quantum domain, such as fidelity, number of qubits, and gate depth. 
In a broader sense, the limited quantum resources should also be utilized efficiently~\cite{herrmann2023quantum}. 
By building on established classical approaches, as well as providing the framework to implement novel solutions, performance can be ensured.
During the \gls{NISQ} era, performance is critical to accommodate the short quantum coherence times.  

\subsection{Scalability} \label{subsec:Scalability}


A primary concern we address with our proposal is scalability.
Scalability itself is a multi-faceted problem, even without including quantum assets.
It contains software scalability, accomodating multiple users, various problem sizes, and flexibility in its solution approaches.
Also, the toolchain does not run in isolation as a single monolithic instance: it co-exists with and leverages surrounding \gls{HPC} software \textit{and} hardware infrastructure.
On the quantum side, almost all vendors employ proprietary interfaces.
High variation in terms of quality and functionality has to be managed correctly.

Besides the differences in software, hardware constraints pose further requirements.
Unique to \gls{QC} is the variety of technologies.
Ensuring smooth \emph{cross-technology} execution is a significant challenge we want to overcome.
A necessary step is the division into hardware-independent and hardware-dependent functionalities.
Still, a generic interface can blend both gradually.
With continuous integration of quantum hardware into classical environments, obstacles of distributed computing have to be resolved.
Also, resource and data management will be relevant for the next generation of \gls{HPCQC}.

As new network topologies arise, communication between nodes will take significant effort.
In \Cref{fig:nodes}, we highlight three modes of operation: \gls{HPC}, quantum, and hybrid components, which need to communicate.
Intercommunication between different modes and inside hybrid nodes poses a massive overhead.
The communication interfaces and protocols require establishing and unifying to resolve such issues.

\subsection{Usability} \label{subsec:Usability}


One of the strengths of a unified toolchain for hybrid \gls{HPCQC} systems is the potential applicability across various fields, such as cryptography~\cite{Q_Crypto}, optimization~\cite{Q_opt}, and quantum simulations~\cite{Q_sim}.
We harness this potential by providing domain experts with a convenient way to plug their unique software tools into a hybrid toolchain and give them control to fine-tune the compilation and execution process.
The inclusion of well-defined interfaces and a well-designed pass manager are essential, as we will explain later. 
Furthermore, \gls{HPC} scientists work continuously on the difficult task of upgrading and maintaining efficient legacy code. 
Hence, such a hybrid \gls{HPCQC} toolchain allows for computational scientists to consider accelerating legacy code with novel quantum modules. 

\subsection{Sustainability} \label{subsec:Sustainability}


The field of QC is rapidly evolving, with new algorithms and techniques emerging and the invention of novel hardware technologies, all while HPC centers build challenging HPCQC integration setups. 
A hybrid toolchain should be adaptive to quantum advances by adopting modular design, forward-compatible data formats for future iterations, and offering a flexible quantum hardware abstraction layer that can be seamlessly expanded as emerging \gls{QPU} technologies are integrated. 

Computational resources, whether classical \gls{HPC} nodes, quantum nodes, or hybrid, are expensive. 
\gls{HPC} scientists try to shift as much computation as possible from the expensive run time to the cheaper compile time. 
Hence, one way to make \gls{QC} more efficient is to allow optimization \gls{AOT}, i.e., during compile time.
Efficiency is gained not only through expressive \glspl{SQIR} but also by using well-developed quantum and classical optimization passes controlled with a pass manager. 

\begin{figure}
    \centering
    \includegraphics[scale=.5]{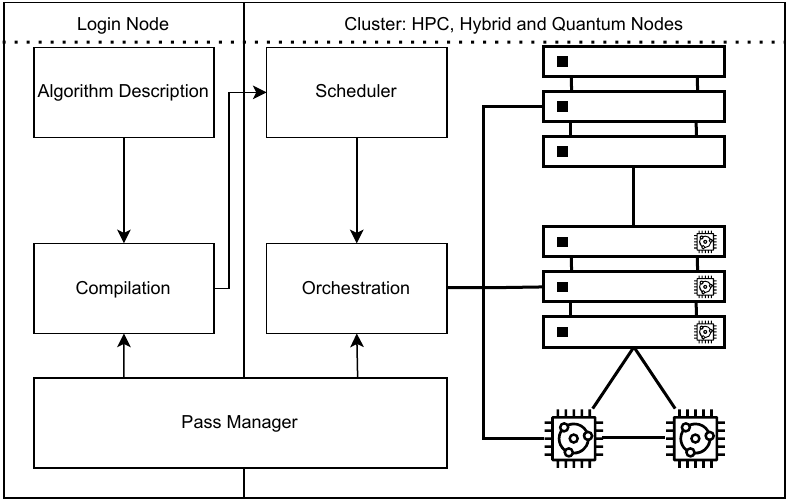}
    \caption{The unified \gls{HPCQC} toolchain from a high-level view as a direct access scenario. The algorithm is described and compiled on the login node. Once the scheduler gets triggered, it assigns the user a portion of \gls{HPC}, Quantum, and Hybrid Nodes. The execution is orchestrated over multiple nodes by the runtime environment.}
    \label{fig:nodes}
\end{figure}

\begin{figure}
    \centering
    \includegraphics[scale=.5]{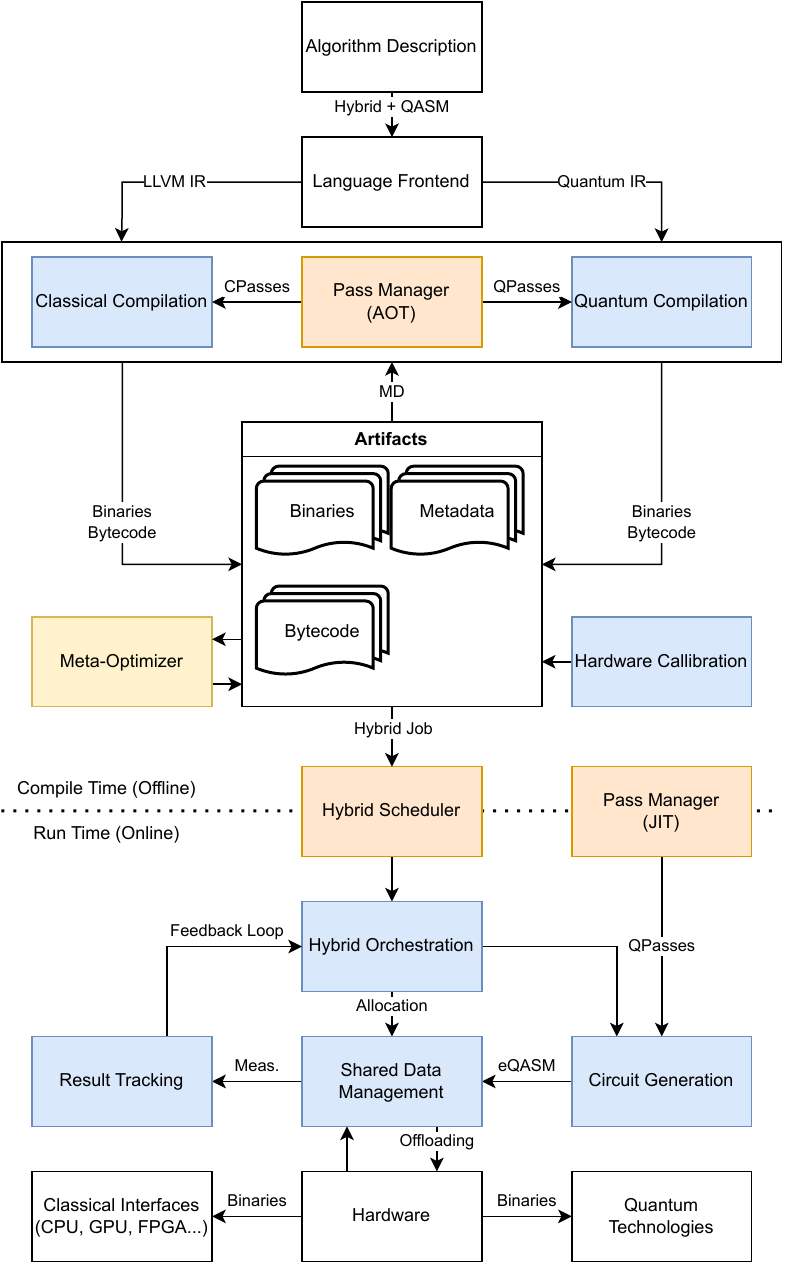}
    \caption{The unified \gls{HPCQC} toolchain from an intermediate-level view. The blue components produce metadata which is consumed by the orange components. The \emph{meta-optimizer} evaluates and uses the generated metadata to control the consumers. Metadata is passed between every component. For simplicity, the annotations have been omitted.}
    \label{fig:toolchain_summary} 
\end{figure}

\section{Proposal: Unified HPCQC Toolchain}\label{sec:Toolchain}

In this section, we present our template of a \emph{unified} \gls{HPCQC} toolchain.
Similar to the standards in formats, it is crucial to define the interaction behavior of individual components.
Standardization is a strict requirement in an \gls{HPC} setting, as all tools have to fit seamlessly to guarantee performance. 
We base our proposal on existing standards to create a platform to interchange tools when needed quickly.
The success of a unified toolchain is based on the adoption in the community and depends on the quality of the available tools.
Our proposal aims at an \gls{HPC} environment that applies to \gls{NISQ} but is also sustainable for fault-tolerant \gls{QC}.

Additionally, it supports different interaction scenarios; \Cref{fig:nodes} depicts the direct access pattern of \gls{HPC}.
A user compiles a program on a login node and submits it to the scheduler.
Cloud interfaces hide the toolchain behind the provider infrastructure for more abstraction.
Specifically, the proposal bridges the gap between user interactions and \gls{HPCQC} hardware.
It contains a \emph{software} tool with generic interfaces on both ends.
It is neither a user frontend nor a hardware controller, and is not tied to a specific execution model.  

Commercial quantum systems offer limited functionality and a restricted execution model.
Multiple integration scenarios are explored by computing centers involving various hardware technologies.
Settings range from loosely coupled provider schemes to on-node quantum acceleration~\cite{HPCQC_review_2023}. 
Concurrently, researchers discover new compilation strategies and introduce programming languages.
It is essential to easily integrate these findings and expose them to users as new functionalities to accelerate the hybrid \gls{HPCQC} development.
The software should not be the bottleneck once major hardware breakthroughs are achieved.
\Cref{fig:toolchain_summary} offers a detailed view of the interactions between components in the toolchain proposal.
It generally describes the flow from a language-independent algorithm formulation through a compilation stage and a runtime stage at which distributed code gets executed.
The components of the proposed toolchain naturally depend on the specific hardware that is used for execution.
For instance, different quantum hardware types differ in their native gate sets, error rates, and qubit connectivity.
The unified toolchain uses abstraction layers to hide hardware details from the user.
Additionally, it contains a data and interfacing model, which is adapted to multiple architectures.
During the workflow, the execution information is progressively exposed and used by this model.

For completeness, we have included a language frontend, which handles hybrid and QASM code, and a hardware calibration component.
The internals of either component exceed the scope of this work and we will not detail them further.
In the following subsections, we comprehensively explain the toolchain, which covers the novel additions to the approach.


\subsection{Stages} \label{subsec:Stages}

As in classical computing, our toolchain acknowledges two distinct stages.
The \emph{compilation stage} happens at compile time, and the \emph{runtime stage} happens at run time.
In \gls{HPC}, this distinction is critical.
A user triggers the compilation stage on a login node, where computation is cheap, and the quantum hardware is not yet known.
Most existing toolchains consider only fully defined quantum programs (kernels), which they generate \gls{AOT}~\cite{mccaskey_xacc_2020}. 
From this point on, most components are working hardware-dependently. 
The final quantum hardware is selected as late as possible, but earliest at this point.
In an arbitrary execution scenario, the full extent of the program might be unknown, and the scheduler selects hardware on the fly.
In these circumstances, a \gls{JIT} compiler transforms the circuit, where the \gls{HPC} system is either idling or used for the process. 
We believe that both compilation modes, \gls{AOT} and \gls{JIT}, should be based on the \emph{same pass manager}, which is detailed in \Cref{sssec:pass_manager}.

\subsection{Producers and Consumers} \label{subsec:Producers_Consumers}

Our design follows the separation of the stages; furthermore, we subdivide into \emph{producers} and \emph{consumers} of metadata.

One single experiment can contain multiple quantum circuits or variants thereof.
Each entire procedure, from compilation to the results, produces a multitude of metadata.
The metadata includes quantum circuit information (e.g., structure and components of the quantum circuit), compilation data (e.g., optimizations applied and error mitigation strategies), execution metadata (e.g., run time and the number of shots), performance metrics (e.g., error rates and circuit depth), as well as system status and environment (e.g., calibration status of the qubits and noise levels).
It can be used to supervise the system continuously and also improves the chances of success.

We aim to update scheduling and \gls{JIT} compilation based on the generated metadata.
Its inclusion adds more depth and complexity to the process, but it is crucial to deviate from fixed routines by adding metadata in the form of \emph{contexts}.
For example, variational algorithms could benefit from this to overcome plateaus or to near convergence.
One can fully optimize performance only with \emph{context-aware} components, which is desirable in \gls{HPCQC}.

\subsection{Components} \label{subsec:Components}
Existing proposals, for example CUDA Quantum~\cite{cuda_quantum}, cover the compilation stage. 
They leverage the well-established LLVM framework to compile hybrid \gls{HPCQC} code efficiently.
This approach is reasonable, but we extend it with a dynamic pass manager, meta-optimizer, and hardware calibration.
There is no established solution for the runtime stage, and we include only the bare minimum. 
We are confident that some objectives are necessary, including ad-hoc circuit generation, data management and offloading, post-processing, and result tracking.
In which form and to what extent this is applicable will be shown by future benchmarks.

\subsubsection{Pass Manager}\label{sssec:pass_manager}

\begin{figure*}
    \centering
    \includegraphics[scale=.5]{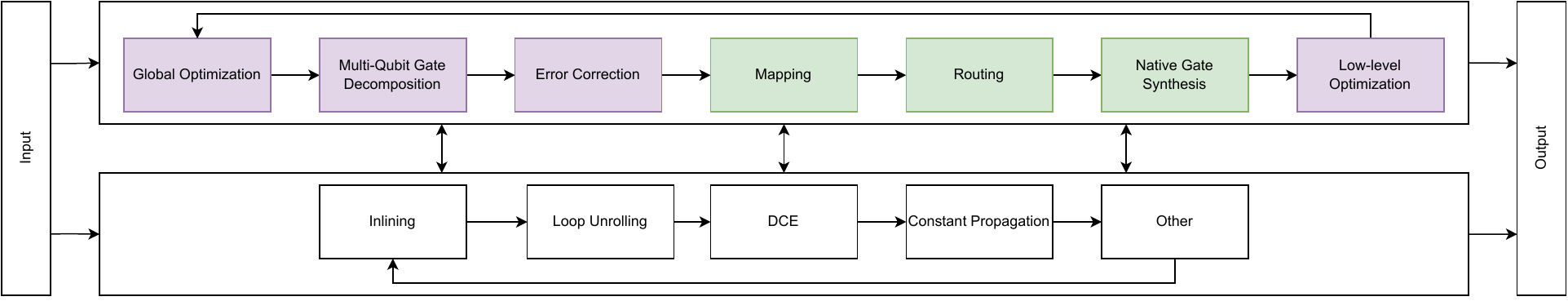}
    \caption{Pass categories utilized by the pass manager. Quantum and classical passes can freely interact. The green steps are mandatory, the purple steps optional. It is possible to repeat or rearrange the steps. Every pass should operate on the same \gls{IR} and have a well defined interface. Implementing this modular structure enables adaptability and  customizability. The chaining of tools has been inspired by t\textbar ket⟩~\cite{sivarajah_tket_2020}.}
    \label{fig:passes} 
    \vspace{-5mm}
\end{figure*}

We extend the concept of compilation passes as used by Qiskit~\cite{qiskit} and LLVM~\cite{LLVM_CGO04} to the runtime stage.
This extension is necessary, assuming future programs build circuits during their execution without relying on predefined kernels.
\Glsfirst{JIT} compilation is an essential part to achieve this vision and already supported by the LLVM compiler Kaleidoscope.
\Cref{fig:passes} shows all quantum passes the pass manager can execute, some of which are optional (purple boxes).
It follows a hybrid compilation process, interweaving classical and quantum optimization routines.
Various techniques transform an input circuit; we have categorized them according to the boxes.
Only the hardware-aware steps are mandatory (green boxes) and can only act once the hardware has been allocated.
However, the compilation is still hybrid during compile and run time.
Classical compilation techniques, like loop unrolling or function inlining, are interleaved with the quantum steps.
Collecting all passes in one component enables unique possibilities and poses a set of new requirements.
An essential concept for this component is reusability.
Existing compilation techniques are not limited to one stage and can be applied multiple times on multiple levels.

Dynamic adaption of circuits is possible based on metadata that becomes available at run time.
E.g., if qubits show subpar fidelity performance, they can be avoided in mappings if the number of available qubits exceeds the demand.
Such context-aware decisions improve the overall quality of service.
Metadata outlives single experiments.
Hence it is available for \gls{AOT} compilation of subsequent experiments.
A similar level of flexibility is also exposed to the user.
Custom compilation sequences can be defined to finetune individual experiments.

All passes must be available in a modular and self-contained package to enable a system to have this level of flexibility.
These packages must clearly define their capabilities, inputs, outputs, and a way to evaluate their performance.
Additionally, since execution is already in an \gls{HPC} environment, passes should leverage the available resources in a parallel fashion.
We advocate using Spack~\cite{Spack}, an open-source package manager, to provide passes to the system.
If passes are self-contained units, they can be managed efficiently by leveraging the utility Spack provides to deploy them as packages.
As an internal representation, we adhere to QIR. 
Its suitability for the runtime stage is still under our review.
It is clear that QIR and, by extension, LLVM are designed for compile time use.
While it is worthwhile to reuse passes, including LLVM as a run time dependency might not be optimal.
Operations at this stage are time-critical; therefore, they could execute distributed, and the overhead of linking LLVM might not be worthwhile.


%


\subsubsection{Meta-Optimizer} \label{sssec:meta_optimizer}

Using machine learning to optimize quantum compilation has been shown to be valid~\cite{Burak_LRZ, Moro2021, quetschlich2023compiler}.
In our toolchain, the meta-optimizer component acts as a global optimization agent.
It improves the complete workflow from selecting and arranging compilation passes, suggesting hardware resources, and recommending decisions on orchestration and data management.
From this \emph{context-aware} approach, we expect to boost the efficiency and performance of hybrid programs.
As for most machine learning systems, the foundation is the available metadata.
Components must provide information about their progress, hence the division into producers and consumers.

\subsubsection{Hybrid Scheduler} \label{sssec:Hybrid_Scheduler}

Classical scheduling is an active area of research.
Adding quantum resources increases the complexity manyfold.
The meta-optimizer can help making informed decisions, but more is needed.
We want to point out some of the challenges subject to our future work.
Our system recognizes two new resource types: \gls{HPCQC} hybrid and pure quantum nodes.
Due to the various quantum technologies, run time decisions must be made based on different properties. 
These may include connectivity restraints, available gate sets and more, based on different available metrics. 
Furthermore, new network topologies have to be supported to accommodate quantum communication.
Existing heuristics can be extended for some of the problems, e.g., resource allocation, utilization, and fair sharing.
One promising idea is to approach quantum resources as a multilevel system, which would allow to schedule quantum systems separately.
State of the art schedulers support new resource types, but depend on a common interface.
These interfaces are implemented to support various layers of abstraction.

\subsubsection{Hybrid Orchestration}  \label{sssec:Orchestration}

We include a component called \emph{hybrid orchestration}, which is part of the runtime environment.
As for the scheduler, the complexity of executing hybrid programs is much higher than the classical counterpart.
The orchestration is responsible for managing heterogeneous loosely coupled components and devise an overall workflow of assigned resources. 
Some conditions must be met to enable tight interaction between components, mainly to avoid performance loss with idle times.
During the execution of a parallel classical problem, (parallel) circuit generation starts, and at the same time, the required resources are allocated and prepared for execution.
For now, we have used eQASM as a placeholder exchange format.
Eventually, different quantum technologies can be utilized similarly to classical cache hierarchies.
Automatic data offloading to the correct levels is inevitable in this situation.
Similarly, the heterogeneous result data has to be consolidated and post-processed.
The main goal behind orchestration is to reduce the system complexity.
Instead of having one omniscient component, the responsibilities are shared over multiple components.

\subsection{Interfaces} \label{sssec:Interfaces}
In this section, we focus on software interfaces; hardware interfaces should be as uniform as possible, but some details can be hidden behind wrappers.
The interfaces define the communication and interaction between different components of the toolchain.
In combination with the artifacts, interfaces provide the basis for the unified toolchain.
They are still under active development as requirements still need to be solidified.
Generally, we have a separation between compile time and run time components which interact through common artifacts.
In \Cref{fig:toolchain_summary}, the interfaces are illustrated with arrows between interacting components.
In the figure, the meta-optimizer interfaces with artifacts, but by extension it is possible to connect it to all producers and consumers.
All components should adhere to a standard interface to allow for scalability and reduce communication overhead.
It clearly defines inputs, outputs, capabilities, errors, and formats.
We strive for a description similar to the \gls{MPI}~\cite{mpi40}, a foundational \gls{HPC} pillar.
The aim is to provide an extensive, yet generic description of how the user can control the interaction between components.
A generic hardware interface should also allow for at-run-time choices for versatile execution.

\subsection{Artifacts}  \label{sssec:Artifacts}
We identify three types of artifacts: \emph{binaries}, \emph{bytecode}, and \emph{metadata}.
\emph{Binaries} contain runnable code; whether quantum, classical, or hybrid.
These programs can be immediately scheduled and executed.
\emph{Bytecode} requires additional processing. 
We assume some parameterized kernels are known ahead of time, such that they can be precompiled.
Under optimistic assumptions about the parameters or the availability of hardware, kernels can be optimized beforehand and finalized at run time with minimal overhead.
In \Cref{fig:toolchain_summary}, both formats lie at the intersection of compile time and run time.
\emph{Metadata} stretches over the complete flow from top to bottom.
Metadata is optional, but highly valuable, as mentioned in \Cref{sssec:meta_optimizer}.
We include hardware calibration under this term as well, as it has the similar function to continuously improve operation.
Metadata must be structured and unified to make it accessible to the user and prepare it for internal use.
To our knowledge, there exists no suitable  format, and one must be established.



\section{Current Status and Future Work} \label{sec:DiscussionandFutureWork}


Currently, two separate groups, both lacking the necessary knowledge, are operating in the quantum domain:
on the one hand, experimental physicists provide the hardware; on the other hand, application engineers use the quantum advantage. 
\Gls{HPC} is the connective link, enabling both to maximize their gains.
Thus, the \gls{HPC} community has to start embracing the quantum advantages. 
This link also opens up access to \gls{QC} for a broader community independent of the existing provider model. 
In the long run, this will also facilitate use cases as described in \Cref{sec:Motivation}.
For complex hybrid algorithms, even small inconsistencies can become a bottleneck when scaling up.
This is reflected in the guiding principles, but can only be explored iteratively.

In the current state, we concentrate our work on immediately applicable solutions, while keeping the long term implications in mind.
The pass manager concept is already valuable for single \gls{QPU} use in the \gls{NISQ} era, hence we started internal development.
We already use a mix of self-developed and external (quantum) compilation techniques, which we use as basis for our evaluation.
Still, we are limited to existing implementations and are open for any contributions.
For the other components, we are still in the conceptual phase and explore prototypes with limited scope. 


An open problem is how to address run time orchestration correctly.  
As we stated before, it is clear which steps must be performed, but not how.
Only through rigorous testing and benchmarking we can establish the requirements.
We plan to do this based on our current work.


The development of the unified \gls{HPCQC} toolchain is an ongoing process. 
The future work will include, but not be limited to, refining the design, testing the implementation on actual hardware offered by the LRZ, and integrating \gls{HPCQC} compilation tools developed within the Munich Quantum Valley (MQV).
Standardization of new interfaces and the extension of existing ones are a continuous process.
While some specifications still need to be prepared for publication, we plan to put them forward in the future.

\section{Conclusion} \label{sec:Conclusion}

We proposed a unified hybrid quantum-classical toolchain.
It is based on emerging standards to facilitate adoption in the community.
Our choices are motivated by \gls{HPC} guiding principles and aim for a generic and customizable solution.
Beyond that, we provide additions to a context-aware, multi-stage pass manager for hybrid compilation.
The meta-optimizer provides this context and also guides scheduling and orchestration.
Interaction between components is primarily clear, but the interfaces need continuous refinement.
Mainly run time components require further research by the community.


\section*{Acknowledgment}
The research is part of the Munich Quantum Valley (MQV), which is supported by the Bavarian state government with funds from the Hightech Agenda Bayern Plus. Moreover, this project is also supported by the Federal Ministry for Economic Affairs and Climate Action on the basis of a decision by the German Bundestag through project QuaST, as well as by the Bavarian Ministry of Economic Affairs, Regional Development and Energy with funds from the Hightech Agenda Bayern. Furthermore, this work is supported by BMW.

\bibliographystyle{IEEEtran}
\bibliography{IEEEabrv, IEEEexample, references} 
\end{document}